\documentclass[12pt]{article}

\usepackage{epsf}
\usepackage{amsmath,amssymb}
\usepackage{latexsym}
\usepackage{graphicx,color}
\usepackage{wrapfig}
\usepackage{latexsym}
\usepackage{graphics}

\usepackage{color}
\usepackage{delarray,color}
\usepackage{fancybox}  

\newcommand {\beq}{\begin{eqnarray}}
\newcommand {\eeq}{\end{eqnarray}}

\newcommand{\be}{\begin{equation}}
\newcommand{\ba}{\begin{eqnarray}}
\newcommand{\ea}{\end{eqnarray}}
\newcommand{\ee}{\end{equation}}

\newcommand{\beqa}{\begin{eqnarray}}
\newcommand{\eeqa}{\end{eqnarray}}
\newcommand{\CR}{\nonumber \\}

\newcommand{\unit}{\hbox to 3.8pt{\hskip1.3pt \vrule height 7.4pt
    width .4pt \hskip.7pt \vrule height 7.85pt width .4pt \kern-2.4pt
    \hrulefill \kern-3pt \raise 3.7pt\hbox{\char'40}}}

\def\matt[#1,#2,#3,#4]{\left(%
\begin{array}{cc} #1 & #2 \\ #3 & #4 \end{array} \right)}

\usepackage{tabularx}

\catcode`\@=11
\@addtoreset{equation}{section}

\catcode`@=12
\relax

\begin{document}

\begin{titlepage}

\setcounter{page}{0}

\renewcommand{\thefootnote}{\fnsymbol{footnote}}

\begin{flushright}
YITP-12-55\\
\end{flushright}

\vskip 1.35cm

\begin{center}
{\Large \bf 
On Supersymmetric Gauge Theories on $S^4 \times S^1$
}

\vskip 1.2cm 

{\normalsize
Seiji Terashima\footnote{terasima(at)yukawa.kyoto-u.ac.jp}
}

\vskip 0.8cm

{ \it
Yukawa Institute for Theoretical Physics, Kyoto University, Kyoto 606-8502, Japan
}

\end{center}

\vspace{12mm}

\centerline{{\bf Abstract}}

We construct supersymmetric 
gauge theory 
on $S^4 \times S^1$.
We find a consistent supersymmetry transformations which
reduced to the 4D $N=2$ supersymmetry transformation studied 
by Pestun by the dimensional reduction on $S^1$.
We find there is no analogue of the usual Yang-Mills action
except in the 4D limit.
We also apply the localization technique 
to the partition function of the theories. 

\end{titlepage}
\newpage

\tableofcontents
\vskip 1.2cm 

\section{Introduction}

The supersymmetric (SUSY) gauge theories on (Euclidean) curved spaces
have been investigated very intensively recently.
A motivation for studying these is 
the possibility of the exact computations of 
the partition function and some BPS operators.
These can be done by the localization technique
in field theories developed by \cite{Nekrasov} \cite{Pestun}.
The work of \cite{Pestun}, where the $N=2$ SUSY gauge theory on 
$S^4$ was considered, 
has been generalized to other geometries
\cite{Kim2}-\cite{HH}.
One of the interesting applications of these 
is the computation of the $N^{\frac{3}{2}}$ scaling of
the partition function of the ABJM model \cite{Marino}.

Especially, the applications of the localization 
technique to the 5D gauge theory on curved space
will be important for studying the still mysterious M5-branes.
Indeed, in the paper \cite{HST} 
the 5d supersymmetric gauge theories on $S^5$ was 
constructed and some interesting results have bee obtained 
for the theories \cite{Ka,SK}.
Since compactifications of the M5-branes give 
varieties of lower dimensional interesting theories,
it will be important to extend the construction of 
the SUSY gauge theory on $S^5$ to other spaces
for studying the M5-branes.

In this paper, we construct supersymmetric gauge theories
on $S^4 \times S^1$.
We find a consistent SUSY transformation which
is reduced to the 4d $N=2$ SUSY transformation studied 
by Pestun \cite{Pestun} by the dimensional reduction on $S^1$.
We find that there is no analog of the usual Yang-Mills action
which does not contain Lorentz violating constant,
except in the 4D limit.
It should be noted that 
we can not use the off-shell 5D supergravity
\cite{KO} to construct a SUSY transformation and actions 
following \cite{Seiberg} because there does not exist a 
field in the supergravity corresponding to the appropriate background 
field which appears in the Killing spinor equation.
Thus we should think the theory has infinite coupling constant.
We apply the localization technique 
to the partition function of the theory on $S^4 \times S^1$ 
following \cite{Pestun} \cite{HH}
and find 
the result is a simple extension of the corresponding partition function 
on $S^4$ with the contributions from Kaluza-Klein modes.
Physical applications of the results in the paper, 
in particular to the M5-branes, are under investigations.

The organization of this paper is as follows:
In section 2 we construct the Killing spinor 
and the consistent SUSY transformations 
of the vectormultiplets and hypermultiplets
for the theory on $S^4 \times S^1$
In section 3, we see that a SUSY Yang-Mills action
for the vectormultiplets on $S^4 \times S^1$
is difficult to construct.
We can construct it only after taking 
the 4D limit by the dimensional reduction on the $S^1$.
In section 4, we apply the localization technique 
to the partition function of the theory on $S^4 \times S^1$ and find
that the result is a simple extension of the corresponding partition function 
on $S^4$.
We conclude with a short discussion in section 5.

\hspace{.5cm}

\noindent
{\bf Note added}: 

As this article neared completion,
we became aware of the very interesting preprint \cite{KK}
where the 5d superconformal index on $S^4 \times S^1$
was calculated and the enhancement of 
global symmetry was checked, however, they just consider the vanishing action
and did not studied possible SUSY actions related to the work of \cite{Pestun}.
The part of their result concerning 
the SUSY transformation for the vectormultiplets
and the localization computations of the partition function
(which is same as the index)
coincide with ours
although we have not introduced the chemical potentials
introduced in \cite{KK}.
They treated the hypermultiplets differently with ours,
however, the partition functions are same.

\section{SUSY transformations}

In this section we will 
construct consistent SUSY transformations 
on $S^4 \times S^1$
by taking a simple ansatz on the Killing spinor.

\subsection{Killing Spinor}

We will construct SUSY gauge theories on $S^4 \times S^1$
following \cite{HST}.
We will use the notations used in \cite{HST} in which 
the theory on $S^5$ was considered.
The indices $m,n,\cdots$ runs from $1$ to $5$,
on the other hand, $\mu,\nu,\cdots$ runs from $1$ to $4$.
As in \cite{HST}, 
we assume the following Killing Spinor equation:
\beq
\nabla_m \xi_I = \Gamma_m \tilde{\xi}_I,
\eeq
where we defined $\nabla_m $ as 
\beqa
\nabla_\mu \xi_I & = & D_\mu \xi_I , \CR
\nabla_5 \xi_I &=& D_5 \xi_I + t_I^{\,\,J} \xi_J.
\eeqa
We further assume
\begin{eqnarray}
 \partial_5 \xi_I=0,
\end{eqnarray}
for simplicity.
The metric of $S^4 \times S^1$ is
\begin{eqnarray}
ds^2_{S^4 \times S^1} &=& (d x^5)^2+
ds^2_{S^4}, \CR
ds^2_{S^4} &=& \ell^2(d\theta^2+\sin^2\theta ds^2_{S^3})
~=~ \frac{dr^2+r^2ds^2_{S^3}}{(1+\frac{r^2}{4\ell^2})^2}
~=~ \frac{\sum_{i=1}^4 dx_i^2}{(1+\frac{r^2}{4\ell^2})^2},
\end{eqnarray}
where 
$r^2=\sum_{i=1}^4 (x^i)^2$ and
$e^a=f\delta^a_n dx^n$ and $f=(1+\frac{r^2}{4\ell^2})^{-1}$.
Here $x^5$ is a coordinate of $S^1$ with radius $R$, thus 
there is an identification $x^5 \sim x^5 + 2 \pi R$.
We can embed the $S^4$ in $R^5$ as $Y_1^2+ \cdots + Y_5^2=l^2$.
The relation between $x^n$ and $Y^i$ ($i=1,\ldots,4$)
is $Y^i=\frac{x^i}{1+\frac{r^2}{4 l^2}}$.

First, $D_\mu \xi_I = \Gamma_\mu \tilde{\xi}_I$
is solved \cite{Pestun} if we take
\begin{eqnarray}
 \xi_I &=& \frac{1}{\sqrt{1+\frac{r^2}{4 l^2}}} \left( 
\epsilon_I+ \frac{x^i\Gamma_i}{2l} \epsilon'_I \right), 
\label{ks} \\
 \tilde{\xi}_I &=& \frac{1}{2l \sqrt{1+\frac{r^2}{4 l^2}}} \left( 
\epsilon'_I- \frac{x^i\Gamma_i}{2l} \epsilon_I \right),
\end{eqnarray}
where $i,j=1,\ldots,4$ which are 4D flat indices and 
$\epsilon_I, \epsilon'_I$ are constants.
For the $S^1$ direction, 
the condition $\nabla_5 \xi_I=t_I^{\,\,J} \xi_J=\Gamma_5 \tilde{\xi_I}$
is written as
\begin{eqnarray}
 \tilde{\xi}_I= t_I^{\,\, J} \Gamma_5 \xi_J.
\end{eqnarray}
Assuming 
\begin{eqnarray}
 t \Gamma_i= \pm \Gamma_i t,
\end{eqnarray}
we have
 \begin{eqnarray}
 (t \epsilon)_I &=& \frac{1}{2l} \Gamma_5 \epsilon'_I \CR
 (t \epsilon')_I &=& \pm \frac{1}{2l} \Gamma_5 \epsilon_I,
\label{ksr}
 \end{eqnarray}
which implies
 \begin{eqnarray}
 (t^2)_I^{\,\, J} = \pm \frac{1}{4l^2} \, \delta_I^{\,\,J},
 \end{eqnarray}
where we assume $[t,\Gamma_5]=0$.
Using (\ref{ksr}), we rewrite (\ref{ks}) as
\begin{eqnarray}
\xi_I &=& \frac{1}{\sqrt{1+\frac{r^2}{4 l^2}}} \left( 
\epsilon_I+ x^i\Gamma_i \Gamma_5 t_I^{\,\, J} \epsilon_J \right). 
\end{eqnarray}

We will later see that
the scaling transformation which enters in the SUSY algebra is
$\rho \sim \epsilon^{IJ} \xi_I \tilde{\eta}_J 
-\epsilon^{IJ} \eta_I \tilde{\xi}_J=
\xi_I t^{IJ} \Gamma_5 \eta_J 
-\eta_I t^{IJ} \Gamma_5 \xi_J $ 
which should vanish for a non-scale invariant theory.
For $t_I^{\,\, I}=0$ which means $t^{IJ}$ is a symmetric tensor for
the $SU(2)_R$, we find $\rho=0$.
Therefore, our conditions essentially fix $t$ 
modulo the $SU(2)_R$ transformation:\footnote{
There is another choice: 
\begin{eqnarray}
 t= i \frac{1}{2l} \sigma_3 \Gamma_5.
\end{eqnarray}
where
\begin{eqnarray}
 \xi_1=&=& \frac{1}{\sqrt{1+\frac{r^2}{4 l^2}}} \left( 
1+ i \frac{x^i\Gamma_i}{2l}  \right) \psi_1, \CR
 \xi_2=&=& \frac{1}{\sqrt{1+\frac{r^2}{4 l^2}}} \left( 
1- i \frac{x^i\Gamma_i}{2l} \right) \psi_2.
\end{eqnarray}
However, in this choice we find that it is difficult to
construct a consistent SUSY transformation
related to this.
Thus, in this paper, we forget this possibility.
}
\begin{eqnarray}
 t_I^{\,\, J}= \frac{1}{2l} (\sigma_3)_I^{\,\, J},
\end{eqnarray}
where
\begin{eqnarray}
 \xi_1=&=& \frac{1}{\sqrt{1+\frac{r^2}{4 l^2}}} \left( 
1+ \frac{x^i\Gamma_i}{2l} \Gamma_5  \right) \psi_1, \CR
 \xi_2=&=& \frac{1}{\sqrt{1+\frac{r^2}{4 l^2}}} \left( 
1- \frac{x^i\Gamma_i}{2l} \Gamma_5  \right) \psi_2.
\end{eqnarray}
This would corresponds to a $SU(2)_R$ gauge field background
in the supergravity, however, the term
$\Gamma_m \tilde{\xi}_I=t_I^J \Gamma_m \Gamma_5 \xi_J$
is absent
in the SUSY transformation of the gravitino of 
the 5D supergravity \cite{KO}.
Thus we can not use \cite{KO} to construct the SUSY invariant action.

\subsection{SUSY transformations}

Now we will construct the SUSY transformations on $S^4 \times S^1$
of the vector multiplets.
First, following \cite{HST} 
we assume that
the SUSY variation of fields on $S^4 \times S^1$ takes the form
\begin{eqnarray}
\delta_\xi A_m &=& i\epsilon^{IJ}\xi_I\Gamma_m\lambda_J~~, \nonumber \\
\delta_\xi\sigma &=& i\epsilon^{IJ}\xi_I\lambda_J~~,\nonumber \\
\delta_\xi\lambda_I &=&
-\frac12\Gamma^{mn}\xi_IF_{mn}+\Gamma^m\xi_ID_m\sigma
+\xi_JD_{KI}\epsilon^{JK}+2\tilde\xi_I\sigma~~, \nonumber \\
\delta_\xi D_{IJ} &=&
-i(\xi_I\Gamma^m \nabla_m \lambda_J+\xi_J\Gamma^m \nabla_m \lambda_I)
+[\sigma,\xi_I\lambda_J+\xi_J\lambda_I]
+i(\tilde\xi_I\lambda_J+\tilde\xi_J\lambda_I)~~ \CR
&=& 
-i(\xi_I\Gamma^m D_m \lambda_J+\xi_J\Gamma^m D_m \lambda_I)
+[\sigma,\xi_I\lambda_J+\xi_J\lambda_I]
+2 i \xi_K t_{IJ} \Gamma_5 \lambda^K,~~
\nonumber\\
\label{susytr}
\end{eqnarray}
where we have used 
\begin{eqnarray}
 s_I^{\,\,K} t_K^{\,\, J} + t_I^{\,\,K} s_K^{\,\, J} 
= -s^{KL} t_{KL} \delta_I^{\,\, J},
\end{eqnarray}
which is valid for an arbitrary symmetric tensor $s_{IJ}$.

Using 
\begin{eqnarray}
 \Gamma^m \nabla_m \tilde{\xi}_I
=\Gamma^m t_I^{\,\,J} \Gamma_5 \Gamma_m \tilde{\xi}_J
=-\frac{3}{4l^2} \xi_I,
\end{eqnarray}
and performing some computations,
we can show that
the commutator of the two
SUSY generators is a sum of a translation ($v^m$), a gauge transformation
($\gamma+i v^m A_m$), 
a dilation ($\rho$), an R-rotation ($R_{IJ}$) and 
a Lorentz rotation ($\Theta^{ab}$):
\begin{eqnarray}
~[\delta_\xi,\delta_\eta]A_m &=& -i v^nF_{nm} + D_m\gamma~~, \nonumber \\
~[\delta_\xi,\delta_\eta]\sigma &=& -i v^nD_n\sigma 
+\rho\sigma~~, \nonumber \\
~[\delta_\xi,\delta_\eta]\lambda_I &=&
-i v^n \nabla_n \lambda_I+i[\gamma,\lambda_I]+\frac32\rho\lambda_I
+{R'}_I^{~J}\lambda_J
+\frac14\Theta^{ab}\Gamma^{ab}\lambda \nonumber \\
 &=&
-i v^n D_n \lambda_I+i[\gamma,\lambda_I]+\frac32\rho\lambda_I
+{R}_I^{~J}\lambda_J
+\frac14\Theta^{ab}\Gamma^{ab}\lambda~~, \nonumber \\
~[\delta_\xi,\delta_\eta]D_{IJ} &=&
-i v^n \nabla_n D_{IJ}+i[\gamma,D_{IJ}]+2\rho D_{IJ}
+{R'}_I^{~K}D_{KJ}+{R'}_J^{~K}D_{IK} \nonumber \\
 &=&
-i v^n D_n D_{IJ}+i[\gamma,D_{IJ}]+2\rho D_{IJ}
+{R}_I^{~K}D_{KJ}+R_J^{~K}D_{IK}
~~,
\label{com}
\end{eqnarray}
where $R_I^{~J}=\epsilon^{JK}R_{IK}$ and
\begin{eqnarray}
v^m &=& 2\epsilon^{IJ}\xi_I\Gamma^m\eta_J~~, \nonumber \\
\gamma &=& -2i\epsilon^{IJ}\xi_I\eta_J\sigma~~, \nonumber \\
\rho &=& -2i\epsilon^{IJ}(\xi_I\tilde\eta_J-\eta_I\tilde\xi_J)=0~~,
\nonumber \\
R_{IJ} &=& -3i(\xi_I\tilde\eta_J+\xi_J\tilde\eta_I
             -\eta_I\tilde\xi_J-\eta_J\tilde\xi_I) 
              -2 i \epsilon^{KL}
             \xi_K \Gamma_5 t_{IJ} \eta_L \nonumber \\
&=&  4 i \epsilon^{KL}
             \xi_K \Gamma_5 t_{IJ} \eta_L ~~, \nonumber \\
\Theta^{ab} &=&
-2i\epsilon^{IJ}(\tilde\xi_I\Gamma^{ab}\eta_J-\tilde\eta_I\Gamma^{ab}\xi_J)~~.
\label{intsym}
\end{eqnarray}
We can see that 
$R_{IJ}=R_{JI}$ and $R_{11}=R_{22}=0$ 
which imply $R_{IJ} \sim t_{IJ}$.
Therefore, the SUSY transformation 
which is inferred from the result of \cite{HST}
is indeed consistent off-shell.


For the hypermultiplets,
we also assume a SUSY transformation of the form given in \cite{HST}:
\begin{eqnarray}\label{hypert}
\delta q_I &=& -2i\xi_I\psi,\nonumber \\
\delta\psi &=&
\epsilon^{IJ}\Gamma^m\xi_I \nabla_m q_J
+i\epsilon^{IJ}\xi_I\sigma q_J
+3 \epsilon^{IJ} \tilde{\xi}_I q_J
+\epsilon^{I' J'}\check\xi_{I'}F_{J'}\nonumber \\
 &=&
\epsilon^{IJ}\Gamma^m\xi_ID_mq_J
+i\epsilon^{IJ}\xi_I\sigma q_J
+2 \epsilon^{IJ} \tilde{\xi}_I q_J
+\epsilon^{I' J'}\check\xi_{I'}F_{J'},\nonumber \\
\delta F_{I'} &=&
2\check\xi_{I'}(i\Gamma^mD_m\psi+\sigma\psi+\epsilon^{KL}\lambda_Kq_L).
\end{eqnarray}
The square of $\delta$ is
\begin{eqnarray}
\delta^2 q_I &=& i v^m D_mq_I -i \gamma q_I
-R_I^{\,\,\, J} q_J
\nonumber \\
\delta^2\psi &=& i v^m D_m\psi-i \gamma\psi
-\frac{1}{4} \Theta^{ab} \Gamma^{ab} \psi
\nonumber \\
\delta^2 F_{I'} &=& i v^m D_m F_{I'}-i \gamma F_{I'}
+{R'}_{I'}^{\,\,\, J'} F_{J'}~~,
\end{eqnarray}
where 
\begin{eqnarray}
v^m &=& \epsilon^{IJ}\xi_I\Gamma^m\xi_J~~, \nonumber \\
\gamma &=& -i\epsilon^{IJ}\xi_I\xi_J\sigma~~, \nonumber \\
R_{IJ} &=& 
2 i (\epsilon^{K L} \xi_K \Gamma^5 t_{IJ} \xi_L) 
~~, \nonumber \\
\Theta^{ab} &=&
-2i\epsilon^{IJ} \tilde\xi_I\Gamma^{ab}\xi_J
~~, \nonumber \\
R'_{I' J'}&=& -2i\check\xi_{I'}\Gamma^mD_m\check\xi_{J'}~~,
\end{eqnarray}
which is consistent with the one for the vectormultiplets.
Therefore, the SUSY transformation for the hypermultiplets
is consistent.

\section{SUSY invariant action}

In this section, we will try to construct 
SUSY invariant actions.
We will see that our SUSY transformation
corresponds to the 4D SUSY transformation of \cite{Pestun}
by the dimensional reduction of $S^1$.
We will drop the total divergent terms below for the notational
convenience.

Now we will try to construct 
a SUSY invariant action for vectormultiplets on $S^4 \times S^1$.
We can show
\begin{eqnarray}
&& \delta_\xi 
\left(
\frac12 F_{mn} F^{mn} - D_m \sigma D^m \sigma -\frac12 D_{IJ} D^{IJ}
+ i \lambda_I \Gamma^m \nabla_m \lambda^I-\lambda_I [\sigma,\lambda^I] 
-i \lambda_I \tilde{\lambda}^I \right) \CR
& = &  i \lambda_I [\Gamma_5 t^{IJ}, \Gamma^{mn}] \xi_J F_{mn}
-2 i D_m \sigma(\xi_I \{ \Gamma^5 t^{IJ},\Gamma^m \} \lambda_J 
-2 i D^{IJ} (\xi_K \Gamma_5 t_{IJ} \lambda^K ) \CR
&& +2 i \xi_I \lambda^I \sigma {\rm Tr}_2 (t^2)
-4i \xi_K \Gamma_5 t^K_J \Gamma_m \Gamma_5 t_I^J \Gamma_m
\lambda^I \sigma 
+{\rm (total \, divergence)},
\end{eqnarray}
and then we find
\begin{eqnarray}
&& \delta_\xi 
\left(
\frac12 F^2 - (D \sigma)^2 -\frac12 D_{IJ} D^{IJ}
+ i \lambda_I \Gamma^m D_m \lambda^I-\lambda_I [\sigma,\lambda^I]
\right. \CR
&& \,\,\,\,\,\,\,\,\,
\left. +2 A_5 t^{IJ} D_{IJ} -t_{IJ} t^{IJ} (6 (A_5)^2-4 \sigma^2)\right) \CR
&  & \,\,\, =  4 i t^{IJ} \left(
(\xi_I \Gamma^m \lambda_J) \partial_5 A_m -(\xi_I \lambda_J) 
\partial_5 \sigma \right).
\end{eqnarray}
This action is not gauge invariant nor SUSY invariant.
However, 
by taking $R \rightarrow 0$, i.e. 
the dimensional reduction to $S^4$,
both problems disappear.
Thus, in this limit to the theory on $S^4$,
we find the invariant action as
\begin{eqnarray}
{\cal L}_{S^4}^{vector} &=& \frac12 F_{mn} F^{mn} - D_m \sigma D^m \sigma
+ i \lambda_I \Gamma^m D_m \lambda^I-\lambda_I [\sigma,\lambda^I] \CR
&&
-\frac12 (D_{IJ}-2 A_5 t_{IJ} )(D^{IJ}-2 A_5 t^{IJ})
-4 t_{IJ} t^{IJ} \left( (A_5)^2-\sigma^2) \right), \CR
\end{eqnarray}
where $\partial_5=0$, which is the usual SUSY Yang-Mills 
Lagrangian of the vector multiplet
used in \cite{Pestun}.
Note that 
the mass terms for the scalars $(A_5, i \sigma)$ are same 
and the auxiliary field 
\begin{eqnarray}
D'_{IJ} \equiv D_{IJ}-2 A_5 t_{IJ}
\end{eqnarray}
transforms as
\begin{eqnarray}
\delta_\xi D'_{IJ} &=&
-i(\xi_I\Gamma^m D_m \lambda_J+\xi_J\Gamma^m D_m \lambda_I)
+[\sigma,\xi_I\lambda_J+\xi_J\lambda_I],
\end{eqnarray}
which is same as the one on $R^4$ and
\begin{eqnarray}
 \delta_\xi\lambda_I &=&
-\frac12\Gamma^{mn}\xi_IF_{mn}+\Gamma^m\xi_ID_m\sigma
+\xi_J D'_{KI}\epsilon^{JK}+2 \left(
\tilde\xi_I\sigma -\Gamma_5 \tilde\xi_I A_5
\right)~~.
\end{eqnarray}
These are consistent with the ones in \cite{Pestun}.

We can construct
a Yang-Mills action on $S^4 \times S^1$ which is invariant under a SUSY
generator, but it is the SUSY exact action which 
will be used as a localization computation and
it depends on constant tensors.
It is difficult to construct the SUSY Yang-Mills action on $S^4 \times S^1$
which reduces to the standard SUSY Yang-Mills action on $S^4$
by the dimensional reduction.
This is partly because the dimensional reduction gives 
a massless scalar in the vector multiplet, but there 
is the mass term for all the 
scalar fields in the SUSY Yang-Mills action 
on $S^4$ by the conformal mapping 
from $R^4$.
To resolve this, we probably need to modify the 
ansatz for the Killing spinor although we will not 
try this in the paper.
We can instead think that the theory has 
infinite gauge coupling constant,
thus the action vanishes.
If the theory with the infinite gauge coupling constant
is at a fixed point of the renormalization group,
this may be justified.
If it is not at a fixed point, 
any quantities will diverge.


Now we will consider the hypermultiplets.
By explicit computations,
we find that the following one is a SUSY invariant Lagrangian 
on $S^4 \times S^1$:
\begin{eqnarray}
\label{hyperla}
{\cal L}_{hyper}&=& \epsilon^{IJ}(D_m\bar q_I D^mq_J-\bar q_I\sigma^2 q_J)
-2(i \bar{\psi} \Gamma^mD_m\psi+\bar\psi\sigma\psi)
\nonumber \\ &&
-i\bar q_ID^{IJ}q_J-4\epsilon^{IJ}\bar\psi\lambda_Iq_J
-\epsilon^{I' J'}\bar F_{I'}F_{J'} \CR
&& -2 t^{IJ} \bar{q}_I D_5 q_J
- 8 t^{KL}t_{KL}\epsilon^{IJ}\bar q_Iq_J
~~.
\end{eqnarray}
Taking the 4D limit, $R \rightarrow 0$, 
we have the SUSY invariant Lagrangian on $S^4$:
\begin{eqnarray}
\label{hyperla4d}
{\cal L}^{hyper}_{S^4}
&=& \epsilon^{IJ}(D_\mu \bar q_I D^\mu q_J
+\bar q_I (A_5)^2 q_J
-\bar q_I\sigma^2 q_J
)
-2(i \bar{\psi} \Gamma^\mu D_\mu \psi
+\bar\psi \Gamma_5 A_5 \psi
+\bar\psi\sigma\psi
)
\nonumber \\ &&
-i\bar q_I {D'}^{IJ}q_J-4\epsilon^{IJ}\bar\psi\lambda_Iq_J
-\epsilon^{I' J'}\bar F_{I'}F_{J'} \CR
&& - 8 t^{KL}t_{KL}\epsilon^{IJ}\bar q_Iq_J
~~.
\end{eqnarray}
This Lagrangian and the SUSY transformation
are different from the ones of \cite{Pestun}.
In \cite{Pestun} the action and the SUSY transformation 
contain quartic terms of the scalars in the hypermultiplets
however, in ours they are quadratic. 
The Lagrangian (\ref{hyperla4d}) will correspond
to the round sphere limit of the Lagrangian in \cite{HH}.
The hypermultiplet Lagrangian in \cite{Pestun} is expected to be 
related to ours by a field redefinition of $D'_{IJ}$
which generate quartic terms.

A mass term for the hypermultiplets is also introduced by
giving a VEV to the vectormultiplets which does not break 
the SUSY.
Here, we can take $\langle A_5 \rangle =m, \sigma=0$ 
and $\langle D_{IJ} \rangle=0$
in the Lagrangian (\ref{hyperla}) or (\ref{hyperla4d}).
Then, a collection of the $m$ dependent terms 
is the mass term.

\section{Localization}

In this section, we apply the localization technique 
to the theory on $S^4 \times S^1$ following\cite{Pestun}.
We take $\xi_I$ as Grassmann-even spinors such that $\delta_\xi$
is a fermionic transformation.


First, we will compute the bilinear of the Killing spinors
which will be used for the localization technique.
For
\begin{eqnarray}
 t_I^{\,\, J}= \frac{1}{2l} (\sigma_3)_I^{\,\, J},
\end{eqnarray}
the explicit form of the Killing spinor is
\begin{eqnarray}
 \xi_1 &=& \frac{1}{\sqrt{1+\frac{r^2}{4 l^2}}} \left( 
1+ \frac{x^i\Gamma_i}{2l} \Gamma_5 \right) \psi_1, \CR
 \xi_2 &=& \frac{1}{\sqrt{1+\frac{r^2}{4 l^2}}} \left( 
1- \frac{x^i\Gamma_i}{2l} \Gamma_5 \right) \psi_2, \CR
 \tilde{\xi}_1 &=& \frac{1}{2l} \Gamma_5 \xi_1, \;\;\;  
 \tilde{\xi}_2 = -\frac{1}{2l} \Gamma_5 \xi_2,
\end{eqnarray}
where 
$\psi_1, \psi_2$ are constant spinors.
Note that we can not impose the $SU(2)$ majorana condition
$\xi_I^\dagger = \epsilon^{JI} \xi_J^T C$
for this.
Instead, we can impose a ``twisted'' $SU(2)$ majorana condition
\begin{eqnarray}
 \xi_I^\dagger = \epsilon^{JI} \xi_J^T C \Gamma_5,
\end{eqnarray}
by imposing $ \psi_I^\dagger = \epsilon^{JI} \psi_J^T C \Gamma_5$.

Now we regard $\xi_I$ as Grassmann even spinors
and will compute 
\begin{eqnarray}
 s &\equiv& \epsilon^{IJ} \xi_I \xi_J, \CR
 v^m &\equiv& \epsilon^{IJ} \xi_I \Gamma^m \xi_J, \CR
 w^{m n}_{IJ} &\equiv&  \xi_I \Gamma^{m n} \xi_J, 
\end{eqnarray}
which appear in $(\delta_\xi)^2$.
We can show that
\begin{eqnarray}
 D_m s &=& 2 t^{IK} w_{m 5 \,IK}, \CR
 D_m v_n &=& 2 t^{IK} \xi_I \Gamma_{nm5} \xi_K
=\epsilon_{mn \mu \nu 5} w^{\mu \nu}_{IK} t^{IK}, \CR
 D_n w_{ml \, IJ} &=& 
= - t_{IJ} \left(
\epsilon_{n ml \mu 5} v^\mu + s(\delta_{5 l} g_{nm}-\delta_{5m} g_{nl})
\right).
\label{e45}
\end{eqnarray}
These implies that $\partial_5 s=\partial_5 v_m=\partial_m v_5=0$
and
\begin{eqnarray}
 D_m v_n+D_n v_m=0,
\end{eqnarray}
i.e. $v^m$ is a Killing vector of $S^4 \times S^1$.

There are choices for the constant spinors $\psi_I$.
In this paper we choose
\begin{eqnarray}
 \Gamma_5 \psi_2=-\psi_2, \;\;\;\; 
\Gamma^{12} \psi_2=\Gamma^{34} \psi_2=i \psi_2,
\end{eqnarray}
because this corresponds to the Killing spinor used
in $S^4$ case \cite{Pestun} as we will see later.
Other choices may be different from this
and interesting to be studied although 
we will concentrate this choice in the paper.
We also normalize the $\psi_i$ as
\begin{eqnarray}
 2 \psi_1^{T} C \psi_2=-1, 
\end{eqnarray}
for the convenience.
Note that this choice is consistent with the twisted $SU(2)$
majorana condition.
Then, we obtain explicitly
\begin{eqnarray}
 s &=& 2 \frac{1}{1+\frac{r^2}{4 l^2} } 
\psi_1^{T} C (1+\Gamma_5 \frac{x^i \Gamma^i}{2l})
(1-\frac{x^i \Gamma^i}{2l} \Gamma_5 ) \psi_2=
-\frac{1-\frac{r^2}{4 l^2}}{1+\frac{r^2}{4 l^2}}=-\cos \theta \CR
 v^5 &=& 1 \CR
 v^\mu \frac{\partial}{\partial x^\mu} &=& 
i \frac{1}{l} \left(
x^1 \partial_2 -x^2 \partial_1 
+x^3 \partial_4 -x^4 \partial_3 
\right) \CR
&=& 
i \frac{1}{l} \left(
Y_1 \frac{\partial}{\partial Y_2} -Y_2 \frac{\partial}{\partial Y_1}
+Y_3\frac{\partial}{\partial Y_4}  -Y_4 \frac{\partial}{\partial Y_3}
\right).
\end{eqnarray}
Note that $v^\mu$ is pure imaginary, but $v^5=1$.
Of course, we can multiply a phase factor to 
the $\psi_I$.
Then, the majorana condition also has an extra phase 
and all bilinears of the Killing spinors are multiplied
by a same phase factor. This can make $v_\mu$ real
although we will not do so.

Let us first concentrate on the vector multiplets.
We take the regulator Lagrangian as  $\delta_\xi V$
where $V = \text{tr}\big[(\delta_\xi \lambda)^\dagger \lambda\big]$. 
Here we recall
\begin{eqnarray}
\delta_\xi\lambda_I &=&
-\frac12\Gamma^{mn}\xi_IF_{mn}+\Gamma^m\xi_ID_m\sigma
+\xi_J D_{I}^{\,\,\,J} +2\tilde\xi_I\sigma~~.
\end{eqnarray}
We define
\begin{eqnarray}
\delta_\xi\lambda_I^\dagger &\equiv&
\frac12 \xi_I^\dagger \Gamma^{mn} F_{mn}-\xi_I^\dagger \Gamma^m D_m\sigma
+\xi_J^\dagger D_{J}^{\,\,\,I} -2\ \xi_K^\dagger \Gamma_5 t_K^{\,\, I} \sigma~~,
\end{eqnarray}
where
we take $\xi_I$ as a twisted $SU(2)$
majorana spinor in order for manifest positive definiteness and 
the condition,  
$\delta_\xi \left(
\int_{S^4 \times S^1} \delta_\xi V\right)=0$.
Note that path-integral contour should be $\sigma^\dagger =- \sigma$
and $D_{IJ}^\dagger=-D^{IJ}$ as in \cite{HST}.

Now we can show that
\begin{eqnarray}
 \delta_\xi\lambda_I^\dagger \,\, \delta_\xi\lambda_I|_{\rm bos} 
&=&
\frac12 F_{mn} F^{mn} - D_m \sigma D^m \sigma 
-\frac12 D_{IJ} D^{IJ}
+2 t_{IJ} t^{IJ} \sigma^2  \CR
&&
-\frac14 s \epsilon^{\mu \nu \rho \sigma} F_{\mu \nu} F_{\rho \sigma}
-\frac12 \epsilon^{\mu \nu \rho \sigma} (D_\rho v_\sigma) F_{\mu \nu} \sigma
\CR
&=&
F_{\mu 5} F^{\mu 5} 
+\frac{1+s}{2} \left( F_- + \frac{\sigma}{1+s} (d v)_- \right)^2
+\frac{1-s}{2} \left( F_+ - \frac{\sigma}{1-s} (d v)_+ \right)^2
\CR
&&
- D_m \sigma D^m \sigma -\frac12 D_{IJ} D^{IJ},
\label{lt1}
\end{eqnarray}
where 
\begin{eqnarray}
 2 (F_\pm)_{\mu \nu} \equiv F_{\mu \nu} \pm \frac12 \epsilon_{\mu \nu \alpha
  \beta} F^{\alpha \beta},
\end{eqnarray}
and $(dv)_{\mu \nu} = D_\mu v_\nu$ which is an antisymmetric
tensor by (\ref{e45}).
Here we have used
\begin{eqnarray}
  (D_\mu v_\nu)^2+ \frac{s}{2} \epsilon^{\mu \nu \alpha \beta}
  (D_\mu v_\nu)   (D_\alpha v_\beta)= 4 \, t_{IJ} t^{IJ} (1-s^2),
 \end{eqnarray}
and other identities follows from the Fierz identities
which are summarized in the Appendix.

The saddle points of (\ref{lt1}) are
\begin{eqnarray}
D_m \sigma=0, \,\,\, D_{IJ}=0, \,\,F_{\mu 5}=0, 
\end{eqnarray}
and for $s \neq \pm1$
\begin{eqnarray}
(F_{\mu \nu})_\pm \mp \frac{\sigma}{2(1 \mp s)}(D_\mu v_\nu)_\pm=0.
 \end{eqnarray}
This implies $D_5 F_{\mu \nu}=0$ and 
\begin{eqnarray}
 \sigma \, d \left( \frac{s}{1-s^2}   
dv + \frac{1}{1-s^2}  *d v \right)=0,
\end{eqnarray}
from the Bianchi identity.
We can check that this implies
\begin{eqnarray}
 \sigma=0, \,\,\, \mbox{and then  } F_{\mu \nu}=0. 
\end{eqnarray}
Only the Wilson line along $x^5$, i.e. 
the constant part of $A_5=a$ in a gauge choice, 
remains as a moduli for $s^2 \neq 1$.
Note that $a$ is taken in the Cartan subalgebra.
The Wilson loop 
\beq
{\rm P} e^{i\int_0^{2 \pi R} dx^5 A_5}=e^{2 \pi i R a} 
\eeq
is invariant under $a \rightarrow  a + \frac{1}{R} H_i$
where $\{ H_i \}$ is any basis of the Cartan algebra
such that the inner product with any weight in any representation 
is an integer.
This means that $a$ is an periodic variable with the above identification.
Except this periodicity, the saddle points for the theory on 
$S^4 \times S^1$
are same as the ones for the theory on the $S^4$.

Next, we consider the localization of the hypermultiplets.
The SUSY transformation for the fermion in the hypermultiplets is 
\begin{eqnarray}
\delta\psi &=&
\epsilon^{IJ}\Gamma^m\xi_ID_mq_J
+i\epsilon^{IJ}\xi_I\sigma q_J
+\epsilon^{I' J'}\check\xi_{I'}F_{J'}
-2 t^{IJ} \Gamma_5 \xi_I q_J,
\end{eqnarray}
For positivity of the action of the hypermultiplets,
we have assumed that
$F$ is ``pure imaginary'' and $q$ is ``real''.
With the rotation of the contours for $\sigma, D_{IJ}, F_{J'}$,
we find
\begin{eqnarray}
(\delta\psi)^\dagger &=&
\xi_I C \Gamma_5 \Gamma^m D_m q_I \Omega
+i \xi_I C \Gamma_5 q^I \Omega \sigma
-\check\xi_{I'}C \Gamma_5 F^{J'}\Omega~~
+2 t^{IJ}\xi_ICq_J\Omega.
\end{eqnarray}
The regulator Lagrangian for the localization will be  $\delta V_{hyper}$
where
\begin{eqnarray}
V_{hyper}&=&  (\delta\psi)^\dagger \psi~~.
\end{eqnarray}
Then, the bosonic part of 
the regulator Lagrangian is 
$\delta V_{hyper}|_{bos}= (\delta\psi)^\dagger \delta \psi$
which becomes
\begin{eqnarray}
\delta V_{hyper}|_{bos}  &=&
\frac12 \epsilon^{IJ} 
D_m \bar{q}_I D_m q_J
+w^{5mn}_{IJ} D_m \tilde{q}^I D_n q^J
-\frac12 \epsilon^{I' J'} \bar{F}_{I'}F_{J'}
-\frac12\epsilon^{IJ}\bar q_I \sigma^2 q_J
\nonumber \\
&& 
- t^{IJ} t_{IJ} \epsilon^{KL} \bar{q}_K q_L
+2 i w^{5 \mu}_{IJ} \bar{q}^I \sigma D_\mu q^J
-2 v^\mu t^{IJ} \bar{q}_I D_\mu q_J
-2 i s t^{IJ} \bar{q}_I \sigma q_J,
\end{eqnarray}
where 
\begin{eqnarray}
 w^{5mn}_{IJ} =\xi_I \Gamma^{5mn} \xi_J.
\end{eqnarray}
In order to derive the saddle point of this,
we note there are
following two inequalities:
\begin{eqnarray}
0 &\leq& | \epsilon^{IJ}\Gamma^m\xi_ID_mq_J|^2
=  \xi_I C \Gamma_5 \Gamma^m D_m q_I \Omega
\epsilon^{IJ}\Gamma^m\xi_ID_mq_J \CR
&&=\frac12 \epsilon^{IJ} 
D_m \bar{q}_I D_m q_J
+w^{5mn}_{IJ} D_m \bar{q}^I D_n q^J,
\end{eqnarray}
and
$ 0\leq \frac12 \epsilon^{IJ} D_m \bar{q}_I D_m q_J$.
Using
\begin{eqnarray}
 w^{5 \mu\nu}_{IJ} D_\mu \bar{q}^I D_\nu q^J
=i \frac12 w^{5 \mu\nu}_{IJ} \bar{q}^I F_{\mu \nu} q^J
-3 t_{IJ} \bar{q}^I D_\nu q^J v^\mu,
\end{eqnarray}
which is valid up to a total divergence term,
we can show 
\begin{eqnarray}
\delta V_{hyper}|_{bos}  & =&
\frac16 \epsilon^{IJ} D_m \bar{q}_I D_m q_J
+\frac53 \left( 
\frac12 \epsilon^{IJ} D_m \bar{q}_I D_m q_J
+w^{5mn}_{IJ} D_m \tilde{q}^I D_n q^J \right)
-\frac12 \epsilon^{I' J'} \bar{F}_{I'}F_{J'}
\nonumber \\
&& 
- t^{IJ} t_{IJ} \epsilon^{KL} \bar{q}_K q_L
\end{eqnarray}
at the saddle points of the vectormultiplets.
Because this is written as a sum of positive definite terms,
we conclude 
\begin{eqnarray}
 q^I=0, \,\,\,\, F^I=0,
\end{eqnarray}
for the hypermultiplets at the saddle points.
Thus, both of the saddle points of vectormultiplets and 
hypermultiplets essentially coincide with the one for $S^4$
\cite{Pestun} \cite{HH}.

In order to compute the 1-loop determinant for the 
regulator Lagrangian, we need to first fix the gauge.
This can be done following \cite{Pestun}.
However, our SUSY transformation for $S^4 \times S^1$
is closer to the one in \cite{HH}. 
Thus, it is more convenient to closely follow \cite{HH}.
As in \cite{HH},
we introduce the BRST transformation $Q_B$
for the field in the vectormultiplets as the usual one
with ghost field $c$ and 
define $Q_B c=i cc +a_0$ where $a_0$ is constant.
We also define
\begin{eqnarray}
 Q c = i \Phi \equiv i(-s \sigma +v^m A_m ),
\end{eqnarray}
where $Q=\delta_\xi$.
We need to introduce other ghosts and their transformation rules:
\begin{eqnarray}
 &&Q_B a_0=Q a_0 =0, \;\;\;\;
 Q_B \bar{c}=B, \,\,  Q \bar{c} =0, \;\;\;\;
 Q_B B=i [a_0,\bar{c}], \,\,  Q B=i v^m \partial_m \bar{c}, \CR 
 && Q_B \bar{a}_0=\bar{c}_0, \,\,  Q \bar{a}_0 =0, \;\;\;\;
 Q_B \bar{c}_0=i [a_0,\bar{a}_0], \,\,  Q \bar{c}_0=0, \CR 
&&  Q_B B_0=c_0, \,\,  Q B_0 =0, \;\;\;\;
 Q_B c_0=i [a_0,B_0], \,\,  Q c_0=0.
\end{eqnarray}
For the gauge fixing, we introduce $\hat{Q}=Q+Q_B$
and take the regulator Lagrangian as
\begin{eqnarray}
 \hat{Q} (V+V_{GF}), 
\end{eqnarray}
where 
\begin{eqnarray}
V_{GF} &=& {\rm tr }(\bar{c} G+\bar{c}B_0+c \bar{a}_0), \CR
 G &= & i D_m A^m + i {\cal L}_v (\Phi- A_5)
=i D_\mu A^\mu 
+  i {\cal L}_{v'} (\Phi- A_5)+ i \partial_5 \Phi,
\end{eqnarray}
and $v'$ is the vector on $S^4$ which is obtained from $v$
by the projection.
This gauge fixing function is taken to be slightly 
different from the one used in \cite{HH},
in order to fix the gauge symmetry related to $x^5$ direction.
The saddle points of the vectormultiplets are unchanged and
other bosonic field vanish at saddle points except
$a_0=A_5$ which is from $\hat{Q} c=0$.

We introduce
\begin{eqnarray}
 \Psi=Q \sigma = i \xi_I \lambda^I, \;\;\; 
\Psi_\mu= Q A_\mu = i \xi_I \Gamma_\mu \lambda^I, \;\;\;
\Xi_{IJ}=\xi_I \Gamma_5 \lambda_J +\xi_J \Gamma_5 \lambda_I,
\end{eqnarray}
which means 
\begin{eqnarray}
 \lambda_I= -i \Gamma_5 \xi_I \Psi - i \Gamma^{\mu 5} \xi_I
  \Psi_\mu +  \xi^J \Xi_{IJ}.
\end{eqnarray}
Then, the fields are classified by boson-fermion and $\hat{Q}$-doublet:
\begin{eqnarray}
&& \,\,\,\,\,  X=(\sigma,A_\mu, \bar{a}_0,B_0), \,\,\,\,\,
 \Xi=(\Xi_{IJ},\bar{c},c) \CR
&&\hat{Q}X=(\Psi,\Psi_\mu+D_\mu c,\bar{c}_0,c_0), \CR
&& \hat{Q} \Xi=(-w_{IJ \rho \sigma} \epsilon^{\mu \nu \rho \sigma}
F_{\mu \nu} +2 w_{IJ}^{5 \mu} D_\mu \sigma- D_{IJ}-2 s t_{IJ} \sigma,B,a_0-\Phi)
\end{eqnarray}
where we have  neglected higher order terms
except ones including moduli $a_0$ and $A_5$.
In terms of these,
we can rewrite $V+V_{GF}$ and then we define $D_{ab}$ by
\begin{eqnarray}
 V+V_{GF}=(\hat{Q}X, \Xi) 
\left( \begin{array}{cc} D_{00} & D_{01} \\ D_{10} & D_{11} 
\end{array} 
\right)
\left( \begin{array}{cc} X \\ \hat{Q} \Xi 
\end{array} \right).
\end{eqnarray}
{}From this equation, we find
\begin{eqnarray}
 \Xi D_{10} X +\Xi D_{11} \hat{Q}\Xi =
\left(
\frac12 w_{IJ \rho \sigma} \epsilon^{\mu \nu \rho \sigma} F_{\mu \nu} 
- w^{5 \mu }_{IJ} D_{\mu } \sigma 
+\frac12 D_{IJ} 
+ t_{IJ} \sigma
\right) \Xi^{IJ} 
\CR
+\frac{i}{4} 
\left( -v_\nu F^{\nu \mu} +v^5 F^{5 \mu} 
-s D^\mu \sigma
+2 w^{5 \mu}_{IJ} t^{IJ} \sigma \right) 
D_\mu c
+\bar{c} G+\bar{c}B_0+c \bar{a}_0,
\end{eqnarray}
%
This $D_{10}$ is different from the one for the round $S^4$ in \cite{HH}
only by the terms $\partial_5 A_\mu$.
Then, the principal symbol of $D_{10}$ is modified.
However, if we 
Fourier-expand $S^1$-direction and
think that the theory is on $S^4$ 
with the Kaluza-Klein towers,
we can see that the principal symbol of $D_{10}$ is same as the one for
$S^4$
because $\partial_5$ is regarded as a constant, not a differential.
Therefore we can apply the index theorem for 
the transversally elliptic operator to each Kaluza-Klein (KK)
momentum.
The $Q^2=(\delta_\xi)^2$ is modified by replacing 
$-i [a_0,*]$ to $D_5 (*)=\partial_5 (*) -i [a_0,*]$
except for $B_o,\bar{a}_0$ which have zero mode only.
Therefore, the 1-loop determinant is 
just given by the product of the one for the vectormultiplets on 
$S^4$ for the KK tower 
with replacement of $a_0 \cdot \alpha$ to
$\frac{n}{R} $ where $\alpha $ is the root of the 
gauge group and the $n$ is the integer KK momentum.

For the hypermultiplets, 
the auxiliary fields can be integrated out trivially.
The $D_{10}$ is modified only by adding term like $D_0 q$
which is not a leading term in the symbol.
Thus, for the hypermultiplets 
the 1-loop determinant is obtained from the 
one for the theory on $S^4$ as for the vectormultiplet.

Finally, 
the instanton contribution at $s=1$, i.e.
at the north pole of $S^4$, is expected to be
the Nekrasov's partition function 
$Z_{inst} (a_0, \epsilon_1=\frac1l,\epsilon_2=-\frac1l,q=1,\beta=R)$
for 5D space \cite{Nekrasov}
because the saddle point equations imply that 
the instantons are localized at the north pole, thus we have the 
quantum mechanics on the instanton moduli space.
There are contributions of the anti-instantons at the south pole
$s=-1$. We will denote the product of these as $|Z_{inst}|^2$.
Note that in the 4D limit, i.e. 
$R \rightarrow 0$, $|Z_{inst}|^2$ is reduced to 
the corresponding term in the one in  \cite{Pestun}
up to an $R$ dependent factor.

Therefore, 
by rescaling $a_0 \rightarrow a_0/l$
our final expression for the 
partition function on $S^4 \times S^1$
obtained from \cite{Pestun} \cite{HH} is
\begin{eqnarray}
\label{pf}
 Z_{S^4 \times S^1} = \int d a_0 |Z_{inst}|^2
\prod_{k \in {\mathbf Z}} 
\frac{
\prod_{\alpha \in \Delta_+} 
\Upsilon \left(
i a_0 \cdot \alpha+i \frac{ l}{R} k \right)
\Upsilon \left(
-i a_0 \cdot \alpha-i \frac{ l}{R} k \right)
}{
\prod_{\rho \in R} 
\Upsilon \left(
i a_0 \cdot \rho+\frac{Q}{2}+i \frac{ l}{R} k \right)
}
\end{eqnarray}
where 
$R$ is the representation of the hypermultiplets
under the gauge group, $\rho$ is a weight of $R$
and $Q=b+\frac1b$ with $b=1$.
Here, according to \cite{HH},
we used the function 
$\Upsilon(x) = \prod_{n_1,n_2 \geq 0} \left(n_1+n_2+x \right)
\left(n_1+n_2+2-x \right)$
to express the regularized infinite product.

The integration variables $a_0$ is periodic,
i.e. $a_0 \sim a_0  +\frac{l}{R} H_i$, and the integrand of 
(\ref{pf}) is indeed periodic under this.
Taking $R \rightarrow 0$ limit with the non-vanishing classical action
which gives non-trivial $q$ dependence,
(\ref{pf}) reduces to the partition function for the theory on $S^4$
\cite{Pestun} \cite{HH} times a numerical factor.
We have introduced the mass term which 
affects the SUSY transformation by $\langle A_5 \rangle$.
This also enters the expression of the partition function
with some shift of the mass \cite{Okuda}.

\section{Conclusion}

In this paper we have constructed supersymmetric gauge theory 
on $S^4 \times S^1$.
We have found there is no analogue of the usual Yang-Mills action
except in the 4D limit.
It should be noted that 
We have applied the localization technique 
to the partition function of the theory on $S^4 \times S^1$ 
following \cite{Pestun} \cite{HH}
and find 
the result is a simple extension of the corresponding partition function 
on $S^4$ with the contributions from Kaluza-Klein modes.

To extend our work to the ellipsoid \cite{HH}
will be straightforward. This will give the result with $b \neq 1$.
In this paper, we only computed the partition function.
Of course, it is interesting to compute the
Wilson loop, the {}'t Hooft loop \cite{Okuda2} and other operators.
(The Wilson loop operators in the maximally SUSY Yang-Mills
theory was considered in \cite{Young})

\section*{Acknowledgments}

I would like to thank 
K. Hosomichi for useful discussions and crucial comments
on many important points on the paper.
I would also like to thank 
Y. Imamura, K. Sakai, Y. Tachikawa and D. Yokoyama for 
useful discussions and
T. Nosaka for 
correcting some errors in the previous version of the paper 
and useful discussions.
The work of S. T. is partly supported by the Japan Ministry of Education,
Culture, Sports, Science and Technology (MEXT), and by the Grant-in-Aid
for the Global COE program ``The Next Generation of Physics, Spun from
Universality and Emergence'' from the MEXT.

\vspace{1cm}

\appendix

\section{Formula for bilinears of Killing spinors}

We can show some relations between the 
bilinears of Killing spinors \cite{HST}.
Here we present them in the form valid for any 5d space.
First, 
\begin{equation}
\Gamma_m\xi_I\cdot v^m ~=~ s \xi_I~~,
\label{key-eq}
\end{equation}
which implies
\begin{eqnarray}
 v_m v^m =s^2.
\end{eqnarray}
Others including $w^{mn}_{IJ}$ are
\begin{eqnarray}
\label{i1}
0 &=& 
-w_{mnIJ} \, w^{mn}_{\,\,\,\,\,\, KL}
+s^2 \left(
\epsilon_{IJ} \epsilon_{KL}+2 \epsilon_{IL} \epsilon_{JK} \right)~~, \\
\label{i2}
0 &=& 2 s \left(
\epsilon_{IJ} \epsilon_{KL}+2 \epsilon_{IL} \epsilon_{JK}
\right) v^p
+2 v_m \left(
\epsilon_{IJ} w^{pm}_{\,\,\,\,\,\,\, KL} - \epsilon_{KL}
w^{pm}_{\,\,\,\,\,\,\, IJ} 
\right) 
\nonumber \\ && 
+ \epsilon^{pmnqr} w_{mn IJ} w_{qrKL}~~, \\
\label{i3}
0 &=& s \left(
8\epsilon_{JK} w^{pq}_{\,\,\,\,\,\,\, LI}
-2\epsilon_{JI} w^{pq}_{\,\,\,\,\,\,\, LK}
-2\epsilon_{LK} w^{pq}_{\,\,\,\,\,\,\, JI} \right)
\nonumber \\ && 
+\epsilon_{JI} v_m \epsilon^{pqmrs} w_{rs LK}
+\epsilon_{LK} v_m \epsilon^{pqmrs} w_{rs JI}
\nonumber \\ && 
-4\left(
w^{qn}_{\,\,\,\,\,\,\, JI} w^{p}_{\,\,\,\, nLK}
-w^{pn}_{\,\,\,\,\,\,\, JI} w^{q}_{\,\,\,\, nLK}
\right)~~.
\end{eqnarray}
These implies
\begin{eqnarray}
\label{i4}
&& 0=v_m w^{pm}_{\,\,\,\,\,\,\, KL}~~, \\
\label{i5}
&& 0=2 s w^{pq}_{\,\,\,\,\,\,\, IJ}
+ v_m \epsilon^{pqrsm}
w_{rs IJ}~~, \\
&& 0= w^{mn}_{\,\,\,\,\,\,\, KI} w_{mn J}^{\,\,\,\,\,\,\,\,\,\,\,\, I}
+3 s^2 \epsilon_{JK}~~, \\
\label{i7}
&& 0=6 \epsilon_{JK} s v^p
-\epsilon^{pmnqr} w_{mn IJ} w_{qr K}^{\,\,\,\,\,\,\,\,\,\,\,\, I}~~, \\
\label{i8}
&& 0= -2 s w^{qp}_{\,\,\,\,\,\,\, JK}
+ \epsilon^{IL} \left(
w^{qn}_{\,\,\,\,\,\,\, IJ} w^{p}_{\,\,\,\, nLK}
+w^{qn}_{\,\,\,\,\,\,\, IK} w^{p}_{\,\,\,\, nLJ}
\right)~~, \\
&& \epsilon^{JL}
(w^{pm}_{~~~~IJ}w^{q}_{~mKL}+w^{qm}_{~~~~IJ}w^{p}_{~mKL})
= \frac32\epsilon_{IK}(s^2 g^{pq}-v^pv^q)\,~~, \\
&& 
\epsilon^{KL} 
w^{pm}_{\,\,\,\,\,\,\, IK} w^{q}_{\,\,\,\, mJL}
=s w^{pq}_{\,\,\,\,\,\,\, IJ}
+\frac{3}{4} \left( s^2 g^{pq} - v^p v^q\right) \epsilon_{IJ}~~.
\end{eqnarray}

\end{document}